\documentstyle[12pt]{article}
\def\bcen{\begin{center}} \def\ecen{\end{center}}
\def\bitem{\begin{itemize}} \def\eitem{\end{itemize}}
\def\btab{\begin{tabular}} \def\etab{\end{tabular}}

\def\IH{{\cal H}}

\def\D{{\cal D}}
\def\hIH{S}
\def\tIH{\tilde{\IH}}

\def\be{\begin{equation}}
\def\ee{\end{equation}}
\def\ba{\begin{eqnarray}}
\def\ea{\end{eqnarray}}
\def\L{{\cal L}}

\def\Re{{\rm Re}}

\def\pullback{\hat{=}}
\def\={\pullback}

\begin{document}

\title{Geometry of Non-expanding Horizons and Their Neighborhoods}

\author{Jerzy Lewandowski}
\maketitle

\centerline{\it Instytut Fizyki Teoretycznej, Uniwersytet Warszawski,
ul. Hoza 69, 00-629 Warszawa, Poland}

\begin{abstract} This is a contribution to MG9 session BHT4.
Certain geometrically distinguished frame on a non-expanding
horizon and in its space-time neighborhood,
as well as the Bondi-like coordinates are constructed.
The construction provides free degrees of freedom, invariants, and
the  existence conditions for a Killing vector field.
The reported results come from the  joint works with Ashtekar
and Beetle \cite{abl}.
\end{abstract}

In the quasi-local
theory of black holes proposed recently by Ashtekar \cite{ABDFKLW} a
BH in equilibrium is described by a 3-dimensional null cylinder $\IH$
generated in space-time by  null geodesic curves intersecting orthogonally
a space-like, 2-dimensional closed surface $\hIH$. The standard
stationarity of space-time requirement is replaced by the assumption
that the cylinder  has zero expansion, that is $\IH$ is
a {\it non-expanding horizon}. This implies, upon the week and the dominant
energy conditions, that the induced on $\IH$ (degenerate)
metric tensor $q$ is Lie dragged by a null, geodesic flow tangent to
$\IH$. The {\it geometry} induced on $\IH$ consists of the  metric
tensor $q$ and the induced covariant derivative $\D$. It is enough for the
mechanics of $\IH$ \cite{ABDFKLW}.
The geometry of a non-expanding horizon is characterized by local
degrees of freedom. They are an arbitrary 2-geometry of the null generators
space $\hIH$,  the rotation scalar, and certain tangential
`radiation' evolving along the horizon.

In the standard, Kerr-Newman case, the event horizon is equipped
with a null Killing vector field. In our general non-expanding horizon
case, however, a Killing vector field may not exits at all.
Our first goal is  a geometric condition which distinguishes a null
vector field $\ell_0$ tangent to $\IH$ and which is satisfied by  the Killing
vector field whenever it exists. We a made  extra assumptions about the stress
energy tensor at $\IH$ that are satisfied for the Maxwell and/or scalar
and/or dylaton fields. The condition distinguishing the null vector
field $\ell_0$ was obtained by making as many components
of the tensor $[\ell, \D]^a_{bc}$ defined on $\IH$ as possible zero,
as we vary $\ell$. But here we give a more geometric definition of
this choice. Due to the evolution equations of $\D$ along $\IH$,
there is a unique extension $(\tIH, \tilde{q}, \tilde{\D})$ of
$(\IH,q,\D)$ in an affine parameter along the null geodesics.
 We claim, that generically $\tIH$
admits a unique global crossection $S_0$ such that its expansion in
the transversal null direction orthogonal to $S_0$ (this information
is contained in  $\tilde{D}$) is zero everywhere on $S_0$. Given the
crossection $S_0$, there is a unique null vector field $\ell_0$
vanishing identically on $S_0$ and such that $\D_{\ell_0} \ell_0 =
\kappa_0 \ell_0$, $\kappa_0\not=0$ being a constant.
Fixing some value $\kappa_0(q,\D)$ determines $\ell_0$ completely.
The shear of $S_0$ vanishes in the null  transversal direction
orthogonal to $\tIH$, iff $\ell_0$ generates a symmetry
of the geometry $(q,\D)$. The commutator
$[\L_{l_0},\D]$ represents the tangential radiation, and  $\IH$ is
not a Killing horizon unless the comutator is zero.

The {\it rotation 1-form potential} $\omega_0$ of $\ell_0$
is  defined by $\D \ell_0\ =\ \omega_0\otimes \ell_0$.
We define a {\it good cut } as a space-like section of $\IH$ such that
the pullback of $\omega_0$ thereon is a harmonic 1-form.
The good cuts define a foliation of $\IH$ invariant with respect
to the flow of $\ell_0$, owing to $\L_{\ell_0}{\omega_0}\ =\
2d\kappa_0\ =\ 0$.

Given $\ell_0$ and the good cuts foliation, we determine
a null frame $(m_0, \bar{m}_0, n_0, \ell_0)$ by using
another null vector $n_0$ orthogonmal to the lives requiring
${n_0}_\mu{\ell_0}^\mu\ =\ -1$, and $\Re{m_0}^\mu K_{,\mu} = 0$
where $K$ is the Gauss curvature of $\IH$, generically
non-constant.

In a neighborhood of  $\IH$, the good cuts foliation and
the distinguished $\ell_0$ define a unique geodesic extension
of the vector field $n_0$. It is used to extend the foliation
and frame to the neighborhood.

The applications and results of this construction are $a)$ invariants
of the horizon and of the neighborhood, $b)$ invariant characterization
and true degrees of freedom of a horizon and of its neighborhood in the
vacuum or Maxwell and/or scalar and/or dylaton case, $c)$
classification of the symmetric isolated horizons, $d)$  necessary and
sufficient conditions for the existence of a Killing vector field, and
the control on the space-times not admitting a Killing vector field.

\section*{Acknowledgments}  This research was  supported in part
by Albert Einstein MPI, CGPG of Pennstate University, and  the Polish
Committee for Scientific Research under  grant no. 2  P03B 060 17.

\end{document}